\documentclass[aip,pop,preprint,showpacs,english,floatfix]{revtex4-1}
\usepackage{babel,rotating,dcolumn}
\usepackage{colordvi,graphicx,color,amsbsy,amsmath,bm,amssymb}
\usepackage{type1cm}
\usepackage{comment}
\usepackage{graphicx}
\usepackage{booktabs}
\usepackage{hhline}

\begin{document}
\title[]
{Anisotropy and intermittency in drift-wave turbulence with zonal flows: a two-dimensional continuous wavelet analysis}

\author{Katsunori Yoshimatsu} 
\email{yoshimatsu@nagoya-u.jp}
\affiliation{Institute of Materials and Systems for Sustainability, Nagoya University, Nagoya, 464-8601, Aichi, Japan}

\author{Zetao Lin} 
\affiliation{Institut de Math\'ematiques de Marseille (I2M), Aix-Marseille Universit\'e, CNRS, 3 place Victor Hugo, 13331 Marseille Cedex 3, France}

\author{Hideaki Miura} 
\affiliation{National Institute for Fusion Science, 
Toki 509-5292, Gifu, Japan}

\author{Kai Schneider} 
\affiliation{Institut de Math\'ematiques de Marseille  (I2M), Aix-Marseille Universit\'e, CNRS, 3~place Victor Hugo, 13331 Marseille Cedex 3, France}
        
\date{\today}
\begin{abstract}
We examine anisotropy and spatial intermittency at small scales in drift-wave turbulence with zonal flows.
We use a two-dimensional directional continuous wavelet transform, which allows simultaneous localization in scale, position, and direction. 
This wavelet analysis is applied to vorticity fields obtained from numerical simulations of the modified Hasegawa--Wakatani model, a reduced model of resistive drift-wave turbulence in magnetized plasmas with zonal flows.
Directional wavelet statistics characterize the anisotropy of the turbulence. 
The second-order moment is enhanced around directions perpendicular to the zonal flow.
Spatial intermittency, characterized by scale-dependent flatness, is more pronounced around directions along the zonal flow.
\end{abstract}

\maketitle

\section{Introduction}
Turbulent dynamics in magnetically confined plasmas are strongly influenced by drift-wave turbulence and zonal flows.
Drift-wave turbulence has a wide range of scales and exhibits intermittency in space and time.
Zonal flows can regulate the turbulence intensity and reduce transport.
In magnetic-confinement fusion, turbulent transport degrades plasma confinement.
For reviews of drift-wave turbulence and zonal flows, see Refs.~\onlinecite{diamond2005zonal,FujisawaReview}.

Hasegawa and Wakatani\cite{HWPRL1983} introduced a reduced two-dimensional (2D) model for resistive drift-wave turbulence in magnetized plasmas, in which density and electrostatic-potential fluctuations are resistively coupled in the presence of a background density gradient.
This model is called the Hasegawa--Wakatani (HW) model.
Numata {\it et al}.\cite{Numata2007} later modified the HW model by removing the zonal component from the resistive coupling term.
Numerical simulations of the modified HW (MHW) model show structures elongated along the zonal flow; see, e.g., Refs.~\onlinecite{Numata2007,Andrew2018,Kadoch22,hnat2025drift,Guillon2025PoP}.
Kadoch {\it et al}.\cite{Kadoch22} reported distinct turbulent flow regions with pronounced zonal flows in the MHW model.

Wavelet analysis provides a useful methodology for turbulent fields.
A wavelet is a function localized in physical space and scale and can also be directional in multiple dimensions.
The coefficients of the continuous wavelet transform (CWT) depend continuously on position and scale, and may also depend on direction in multiple dimensions.
Orthogonal wavelet transforms have efficient fast algorithms, while their positions, scales, and directions are discrete and constrained by the multiresolution construction.
By contrast, Fourier analysis represents turbulent fields by modes localized in wavenumber space but not in physical space.
Reviews of wavelet applications to turbulence, including plasma turbulence, are given in, e.g., Refs.~\onlinecite{FargeARF,KaiARF2010,FargeSchneider2015,FargeArX}.
Continuous wavelet analysis has also been applied to experimental plasma data; see, e.g., Ref.~\onlinecite{FujisawaReview}.
An early application introduced the wavelet bicoherence to detect short-lived intermittent nonlinear coupling.~\cite{MilligenPRL}
Continuous wavelet analysis has also been applied to observational and numerical simulation data of astrophysical plasma turbulence to examine relationships among large-amplitude structures, background fluctuations, and kinetic Alfv\'en wave features.\cite{KaiPRX}
Orthogonal wavelet analysis has been used to extract coherent structures.\cite{FargeSchneider2015} 
For example, coherent bursts were extracted from turbulent edge plasmas in magnetic fusion devices.~\cite{farge2006extraction} 

Spatial intermittency implies that, at a given time, the intensity of turbulent activity varies strongly, with intense activity occurring irregularly.
In hydrodynamic turbulence, the scale-dependence of the spatial localization of intense fluctuations has been studied (e.g., Refs.~\onlinecite{farge90CTR,Meneveau1991}).
Temporal intermittency was reported in experimental studies dating back to the late 1940s,\cite{Townsend1949,BT1949}
and examined at different scales using band-pass-filtered signals.\cite{KennedyCorrsin1961}

In this work, we examine the anisotropy and spatial intermittency
at small scales
in drift-wave turbulence with zonal flows.
We analyze vorticity fields obtained from numerical simulations using wavelet statistics based on the CWT.
A 2D complex-valued directional Morlet wavelet is used.

The remainder of this manuscript is organized as follows. 
In Sec.~\ref{sec:DWT}, we present the MHW model and the simulation data. 
Section~\ref{sec:CWT} briefly reviews the 2D CWT and introduces the scale-, position-, and direction-dependent statistical quantities. 
Section~\ref{sec:RES} presents the numerical results of wavelet statistics for the different flow cases. 
Finally, Sec.~\ref{sec:concl} gives the conclusions.

\section{Drift-wave turbulence with zonal flows}\label{sec:DWT}
We present the MHW model and describe the numerical simulations.

\subsection{Modified Hasegawa--Wakatani model} \label{subsec:HW}
We consider drift-wave turbulence in a 2D slab geometry with a uniform background density gradient.
The MHW model\cite{Numata2007} is employed as a reduced fluid model.
We use Cartesian coordinates $(x,y)$ in the slab plane, corresponding to the radial and poloidal directions, respectively.
We assume a uniform magnetic field directed perpendicular to the $x-y$ plane.
Ions are treated as cold.
The dimensionless MHW equations for the vorticity $\omega$ and density fluctuation field $n$ in the doubly periodic domain $\Omega=[0,L]^2$ are given by
\begin{eqnarray}
\frac{\partial \omega}{\partial t} + \left[\phi,\omega \right] &=& c ({\widetilde{\phi}} - {\widetilde{n}}) + \nu \nabla^2  \omega, \label{MHWome} \\
\frac{\partial n}{\partial t} + \left[\phi, n \right] &=&  c ({\widetilde{\phi}} - {\widetilde{n}}) +D \nabla^2  n - \Gamma \partial_y \phi.\label{MHWden}
\end{eqnarray}
Here, 
 $\phi(x,y,t)$ is the electrostatic potential, $\omega = \nabla^2 \phi$, $\nabla = (\partial/\partial x, \partial/\partial y)$, $t$ is time, and 
$[g,h]=\partial_x g\,\partial_y h-\partial_y g\,\partial_x h$ denotes the Poisson bracket. 
Arguments such as $x$, $y$ and $t$ are omitted for brevity, unless otherwise stated.
The non-zonal component of a field $g$ is defined as $\widetilde g = g-\bar g$, 
where ${\bar g} = L^{-1}\int_0^L g(x,y)\,dy$ is the zonal component, i.e., the average over $y$ at fixed $x$.
The resistive coupling terms in Eqs. (\ref{MHWome}) and (\ref{MHWden}) are constructed so that they do not act on zonal components, in contrast to the HW formulation.\cite{HWPRL1983}
Details of the normalization and background density profile are given in Ref.~\onlinecite{Numata2007}.
The term $-\Gamma \partial_y \phi$ is due to the background density gradient,
where $\Gamma$ is the density-gradient parameter.
By rescaling the variables and parameters as $t'=\Gamma t$, $\phi'=\phi/\Gamma$, $n' = n/\Gamma$, $\omega'=\omega/\Gamma$, $c'=c/\Gamma$, $\nu'=\nu/\Gamma$, and $D'=D/\Gamma$,
Eqs.~(\ref{MHWome}) and (\ref{MHWden}) retain the same form with $\Gamma=1$.
Hereafter, the primes are omitted.
The parameters $\nu$ and $D$ denote
the coefficients of the viscous and diffusive terms, respectively, 
while $c$ is the adiabaticity parameter.
As in Ref.~\onlinecite{Kadoch22}, 
we use standard Laplacian viscous and diffusive terms rather than the hyperviscous and hyperdiffusive terms used in Ref.~\onlinecite{Numata2007}.
The velocity is given by $(u_x,u_y)=(-\partial \phi/\partial y, \partial \phi/\partial x)$. 

\subsection{Numerical simulations} \label{subsec:DNS}
\begin{figure}[bt!]
\begin{center}
\includegraphics[width=7cm,keepaspectratio]{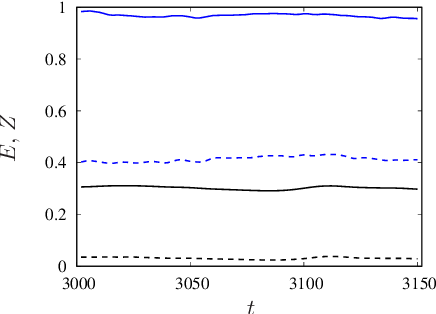}
\end{center}
\caption{Time series of the spatially averaged kinetic energy $E(t)$ and enstrophy $Z(t)$ over the period used for averaging. 
Solid and dashed curves represent $E(t)$ and $Z(t)$, respectively; black and blue curves denote the cases c4 and c07, respectively. Here, $Z(t)$ is the spatial average of $\omega^2/2$.} \label{figEZ}
\end{figure}
\begin{table}[tbp]
  \centering
  \caption{Statistical quantities of the flow. The Reynolds number is defined by $ R_\lambda = \lambda E_{\mathrm{rms}}^{1/2} /\nu $, where $\lambda=(E_{\mathrm{rms}}/Z_{\mathrm{rms}})^{1/2} $. 
  Here, $E_{\mathrm{rms}}$ is defined in the same way as $Z_{\mathrm{rms}}$, with $E(t)$ replacing $Z(t)$.}  \label{tab:stats}
  \begin{tabular}{cccccc}
    \hhline{=====}
    Runs & $c$  &  \quad $\nu$ & \quad  $ R_\lambda $ &  \quad$t_d/t_k$ \\
    \hline
      c4 & $4.0$ &  \quad $ 5.0 \times 10^{-3}$ & \quad 343 & \quad  37.3 \\
    \hline
     c07 & $0.7$ &  \quad $ 2.5 \times 10^{-3}$  & \quad 602  & \quad 136 \\
    \hhline{=====}
  \end{tabular}
\end{table}
\begin{figure}[bt!]
\begin{center}
\includegraphics[width=16cm,keepaspectratio]{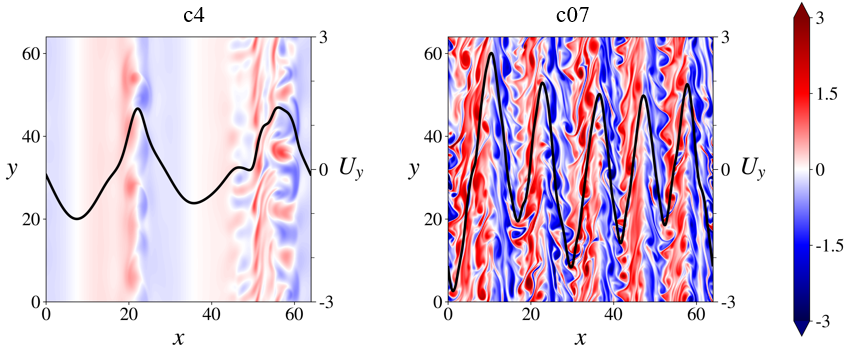}
\end{center}
\caption{Visualization of the vorticity field $\omega$ at $t=t_s=3076$.
The solid curves represent the zonal-flow velocity $U_y$.} \label{viFLD}
\end{figure}
We use two datasets from numerical simulations of the MHW model.
The turbulent fields are governed by Eqs.~(\ref{MHWome}) and (\ref{MHWden}) in the doubly periodic domain $\Omega=[0,L]^2$.
The domain size is set to $L=64$.

One dataset corresponds to $c=4$ and was obtained by Kadoch {\it et al.}\cite{Kadoch22}.
The other corresponds to $c=0.7$ and was obtained in this work.
These cases are labeled c4 and c07, respectively.
The parameters $\nu$ and $D$ are set to $\nu=D=5\times10^{-3}$ for c4 and $\nu=D=2.5\times10^{-3}$ for c07.
The c07 case is chosen as a representative state that retains zonal-flow structures and exhibits stronger small-scale activity than the c4 case.
Both datasets were computed with the numerical method used in Ref.~\onlinecite{Kadoch22}.
The equations are solved using a Fourier pseudo-spectral method and a fourth-order Runge--Kutta method.
The two-thirds rule is used for dealiasing.
The number of grid points is $1024^2$, and the time increment is $\Delta t=5\times10^{-4}$.
The initial fields at $t=0$ are Gaussian random fields.
The spatial averages of $\omega$ and $n$ are zero throughout the simulations.

Figure~\ref{figEZ} shows the time series of the spatially averaged kinetic energy $E(t)$ and enstrophy $Z(t)$ over the period used for averaging.
For both c4 and c07, $E(t)$ and $Z(t)$ remain quasi-stationary over this period.
Table~\ref{tab:stats} summarizes several statistical quantities.
These statistics are obtained from 75 snapshots sampled every 2 time units.
The averaging period is denoted by $t_d$, with $t_d=148$.
The eddy turnover time $t_k$ is defined as $(2Z_{\mathrm{rms}})^{-1/2}$, where $Z_{\mathrm{rms}}$ denotes the root-mean-square of $Z(t)$ over this period.

Figure~\ref{viFLD} shows visualizations of the vorticity field $\omega$ at time $t_s$ within this period for c4 and c07.
The zonal-flow velocity $U_y$, defined by $U_y(x,t)=L^{-1}\int_0^L u_y(x,y,t)\,dy$, is superimposed on each visualization.
In both cases, turbulent stripe-like structures are seen, suggesting the coexistence of zonal flows and turbulence.
It can be seen that, for c4, the region with $U_y>0$ is turbulent, whereas the region with $U_y<0$ is much less turbulent.
For c07, the flow appears more turbulent than for c4 throughout the domain.
Related flow visualizations were also reported, e.g., in Refs.~\onlinecite{Andrew2018,Kadoch22,hnat2025drift,Guillon2025PoP}.

\section{2D Continuous Wavelet Analysis} \label{sec:CWT}
We briefly summarize the 2D CWT with a directional wavelet, 
which provides simultaneous localization in scale, position, and direction. 
Readers interested in details of the CWT may refer to textbooks, e.g., Refs. \onlinecite{daubechies1992,antoine2004}.
We then introduce wavelet statistics that characterize the scale-dependence of anisotropy and spatial intermittency.
\subsection{2D complex continuous wavelet transform}\label{sec2dCWT}
The CWT is constructed from a mother wavelet $\psi(\bm{x}) \in L^2(\mathbb{R}^2)$, where $\bm{x}=(x,y)\in\mathbb{R}^2$.
The mother wavelet satisfies the zero-mean condition
\begin{equation}
\int_{\mathbb{R}^2}\psi(\bm{x})\,d\bm{x}=0,
\end{equation}
and the admissibility condition $\int_{\mathbb{R}^2}k^{-2}\,|\hat{\psi}(\bm{k})|^2\,d\bm{k}<\infty $,
where $\hat{\psi}(\bm{k})$ is the Fourier transform of $\psi(\bm{x})$ defined by
${\hat \psi}({\bm k}) = (2\pi)^{-2} \int_{\mathbb{R}^2} \psi({\bm x}) \exp (-i {\bm k} {\bm \cdot} {\bm x})d {\bm x}. $
Here, ${\bm k}$ is the wave vector and $k=|{\bm k}|$.

We consider a complex-valued directional wavelet. 
By dilation, translation, and rotation,
we obtain a family of wavelets at scale $a$, position $\bm{X}$, and direction $\Theta$:
\begin{equation}
\psi_{a,\bm{X},\Theta}(\bm{x}) = \frac{1}{a}\psi\!\left[ R^{-1}(\Theta)\frac{\bm{x}-\bm{X}}{a} \right].
\end{equation}
Here, $a\in\mathbb{R}^+$, $\bm{X}\in\mathbb{R}^2$, and $R(\Theta)$ is a 2D rotation matrix,
\begin{equation}
R(\Theta)=
\begin{pmatrix}
\cos\Theta & -\sin\Theta\\
\sin\Theta & \cos\Theta
\end{pmatrix},
\end{equation}
where the angle $\Theta$ is measured counterclockwise from the positive $x$-axis.
The functions $\psi_{a,\bm{X},\Theta}(\bm{x})$ are well-localized in scale and position, and are associated with direction $\Theta$.
They also satisfy the zero-mean condition.
For a scalar field $f\in L^2(\mathbb{R}^2)$, the 2D CWT is defined by
\begin{equation}
W[f](a,\bm{X},\Theta) = \int_{\mathbb{R}^2} f(\bm{x})\,\psi^{*}_{a,\bm{X},\Theta}(\bm{x})\,d\bm{x}, 
\end{equation}
where the superscript $^*$ denotes the complex conjugate.

In this study, we work in a doubly periodic domain $\Omega=[0,L]^2$.
The field $f(\bm{x})$ is assumed to be $L$-periodic in both $x$ and $y$.
The periodized form of $\psi_{a,\bm{X},\Theta}(\bm{x})$ is defined by
\begin{equation}
\psi^{\mathrm{per}}_{a,\bm{X},\Theta}(\bm{x}) =\sum_{\bm{\ell}\in\mathbb{Z}^2} \psi_{a,\bm{X},\Theta}(\bm{x}+L\bm{\ell}).
\end{equation}
Then, if we identify $f$ with its periodic extension to $\mathbb{R}^2$, the CWT can be written equivalently as
\begin{equation}
\int_{\mathbb{R}^2}
f(\bm{x})\,\psi^{*}_{a,\bm{X},\Theta}(\bm{x})\,d\bm{x}
= \int_{\Omega}f(\bm{x})\,\psi^{\mathrm{per},*}_{a,\bm{X},\Theta}(\bm{x})\,d\bm{x}.
\end{equation}
We define the normalized wavelet coefficient by
\begin{equation}
W^{\mathrm{per}}[f](a,\bm{X},\Theta) =\frac{1}{L^2} \int_{\Omega} f(\bm{x})\,\psi^{\mathrm{per},*}_{a,\bm{X},\Theta}(\bm{x})\,d\bm{x}.\label{cwt2dP}
\end{equation}

We use a 2D complex directional Morlet wavelet,
whose basic form is $\exp \left({i {\bm k}_\psi {\bm \cdot}  {\bm x} } -\left| {\bm x} \right|^2/2\right) $.
We take ${\bm k}_\psi =(k_\psi,0)$ without loss of generality.
Here, $k_\psi$ denotes the dimensionless central wavenumber.
We chose $k_\psi=6$ following the standard choice in, e.g., Ref.~\onlinecite{FargeARF}. 
The scale $a$ is approximately related to the wavenumber $k$ by 
\begin{equation}
a k  \simeq k_\psi . \label{ak}
\end{equation}
See, e.g., Ref. \onlinecite{Jori}.
For a given wavenumber $k$, the scale $a$ is adjusted according to Eq.~(\ref{ak}) when $k_\psi$ is varied.
At fixed $k$, a larger $k_\psi$ corresponds to a larger $a$.
The wavelet is then broader in physical space, while its bandwidths with respect to wavenumber magnitude and angle are narrower.
A smaller $k_\psi$ has the opposite effect.

Wavelet coefficients at each scale and direction can be computed efficiently for doubly periodic data by an inverse fast Fourier transform (see, e.g., Refs.~\onlinecite{FargeARF,Jori}).
For the periodized Morlet wavelet used here, the zero Fourier mode is set to zero in the numerical implementation.
FFTE\cite{FFTE} was used to compute the Fourier transforms in the CWT analysis.
The angle $\Theta$ corresponds to the direction of the rotated central wave vector $R(\Theta){\bm k}_\psi$.
With this wavelet, the coefficient of a real-valued field $f$ satisfies
$W^{\mathrm{per}}[f](a,{\bm X},\Theta) = W^{\mathrm{per},*}[f](a,{\bm X},\Theta+\pi)  $. 
Therefore, for the modulus of the wavelet coefficient and its powers, it is sufficient to consider $0 \le \Theta < \pi$.
Hereafter, we omit the superscript ``per'' and write $W[f](a,\bm{X},\Theta)$.

\subsection{Wavelet statistics}\label{secWS}
We focus on the vorticity $\omega$, since it exhibits pronounced small-scale fluctuations.
We define the second- and fourth-order moments of the wavelet coefficient magnitude as
\begin{equation}
M_\ell(k,\Theta)=\left\langle |W[\omega](k,\bm{X},\Theta)|^\ell \right\rangle, \quad \ell=2,4,
\end{equation}
where $\langle \cdot \rangle$ denotes the average over space ($\bm{X}\in\Omega$) and time.
Using Eq.~(\ref{ak}), we express the scale $a$ in terms of the wavenumber $k$.
The angular averages are defined by
\begin{equation}
M_\ell^\Theta(k) = \left\langle M_\ell(k,\Theta)\right\rangle_\Theta, \quad \ell=2,4,
\end{equation}
where $\langle \cdot \rangle_\Theta$ denotes the average over $\Theta$.
The quantity $|W[\omega](k,\bm{X},\Theta)|^2/a^2$ denotes the local wavelet spectrum at a given time.~\cite{FargeARF}
Farge {\it et al}.~\cite{farge90CTR} introduced the local intermittency measure (LIM).
Here, we extend it to a direction-dependent form,
\begin{equation}
I(k,\bm{X},\Theta) = \frac{|W[\omega](k,\bm{X},\Theta)|^2}{M_2(k,\Theta)}, \label{LIM}
\end{equation}
which characterizes the spatial distribution of the local wavelet spectrum for each wavenumber and direction.
If the local wavelet spectrum is spatially uniform and nonzero at fixed $k$ and $\Theta$, $I(k,\bm{X},\Theta)=1$ for all $\bm{X}$.

We next define the global wavelet spectrum, the wavelet flatness, and the anisotropy measure.
Following Ref.~\onlinecite{FargeARF}, we introduce the global wavelet spectrum by
\begin{equation}
{\mathcal Z}(k,\Theta)=\frac{M_2(k,\Theta)}{a^2}.\label{WSpe}
\end{equation}
This quantity is related to the Fourier spectrum through the Fourier transform of the wavelet.\cite{FargeARF} 

The scale-dependent anisotropy measure is defined as
\begin{equation}
A^{\mathcal Z}(k,\Theta)=\frac{{\mathcal Z}(k,\Theta)}{{\mathcal Z}^\Theta(k)},  \label{AniE}
\end{equation}
where
\begin{equation}
{\mathcal Z}^\Theta(k)=\langle {\mathcal Z}(k,\Theta)\rangle_\Theta .\label{AveWSpe}
\end{equation}
Under the assumption of statistical isotropy, $A^{\mathcal Z}(k,\Theta)=1$ for all $\Theta$.\cite{FargeARF}

Wavelet flatness has been used to characterize scale-dependent spatial intermittency (see, e.g., Refs.~\onlinecite{Meneveau1991,Camussi2021}).
Here, we define the wavelet flatness at each angle $\Theta$ as
\begin{equation}
F(k,\Theta) = \frac{M_4(k,\Theta)}{\left[M_2(k,\Theta)\right]^2}. \label{WFlat}
\end{equation}
For comparison, if a complex wavelet coefficient is a zero-mean Gaussian random variable, the Gaussian fourth-moment relations for the real and imaginary parts of the wavelet coefficient give $2\le F(k,\Theta)\le 3$. 
If these two parts have equal variances and are uncorrelated, then $F(k,\Theta)=2$.
From Eq.~(\ref{LIM}), this flatness is the second moment of the LIM:
\begin{equation}
F(k,\Theta)=\langle [I(k,\bm{X},\Theta)]^2\rangle.\label{FlatLIM}
\end{equation}
The anisotropy measure for wavelet flatness is then defined by
\begin{equation}
A^F(k,\Theta) = \frac{F(k,\Theta)}{F^\Theta(k)},\label{AniF}
\end{equation}
where
\begin{equation}
F^\Theta(k) = \frac{M_4^\Theta(k)}{\left[M_2^\Theta(k)\right]^2}. \label{aveFlat}
\end{equation}
The measure $A^F(k,\Theta)$ quantifies the scale-dependent anisotropy of spatial intermittency.
Under the assumption of statistical isotropy, $A^F(k,\Theta)=1$ for all $\Theta$.

\section{Wavelet-Based Numerical Results } \label{sec:RES}
We consider the two cases c4 and c07, whose vorticity fields $\omega$ have distinct spatial structures.
Each field is analyzed in terms of scale, position, and direction using the wavelet transform.
Local wavelet spectra and the LIM are presented first.
Wavelet statistics of $\omega$ are then examined.
The 2D directional Morlet wavelet is used with the central wavenumber $k_\psi=6$. 
The angle $\Theta$ is measured counterclockwise from the positive $x$-axis and specifies the direction of the rotated central wave vector $R(\Theta){\bm k}_\psi$.
Thus, the wavelet directions at $\Theta=0$ and $\pi/2$ are parallel to the mean density gradient and the zonal flow, respectively.
Since the statistics below are based on the modulus of the wavelet coefficient, it is sufficient to consider $0\le\Theta<\pi$.
\subsection{Visualization in wavelet space} \label{Viswave}
\begin{figure}[bt!]
\begin{center}
\includegraphics[width=15cm,keepaspectratio]{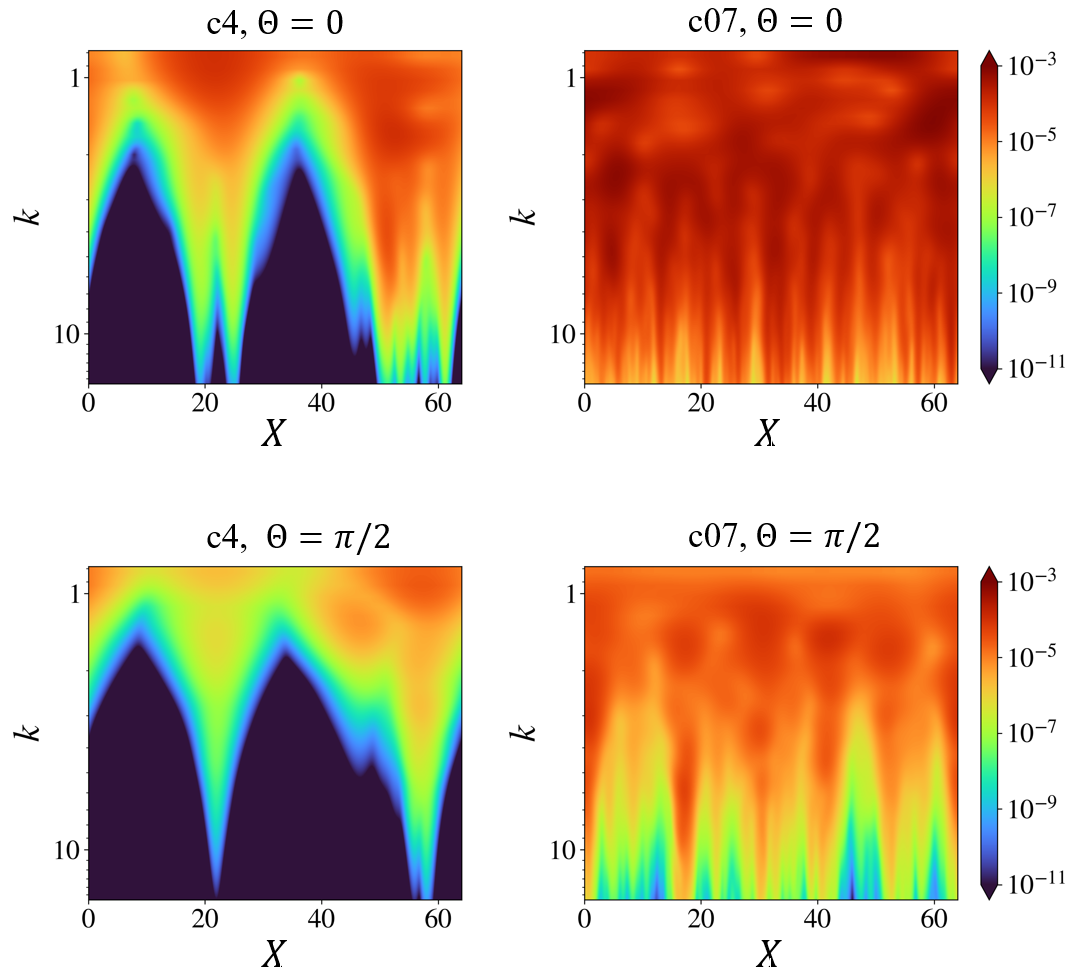}
\end{center}
\caption{Scalograms of $\omega$ at $t=t_s$, obtained from the $Y$-averaged local wavelet spectrum $\langle |W[\omega](k,{\bm X},\Theta)|^2 \rangle_Y /a^2$.} \label{visW1D}
\end{figure}
\begin{figure}[th!]
\begin{center}
\includegraphics[width=15cm,keepaspectratio]{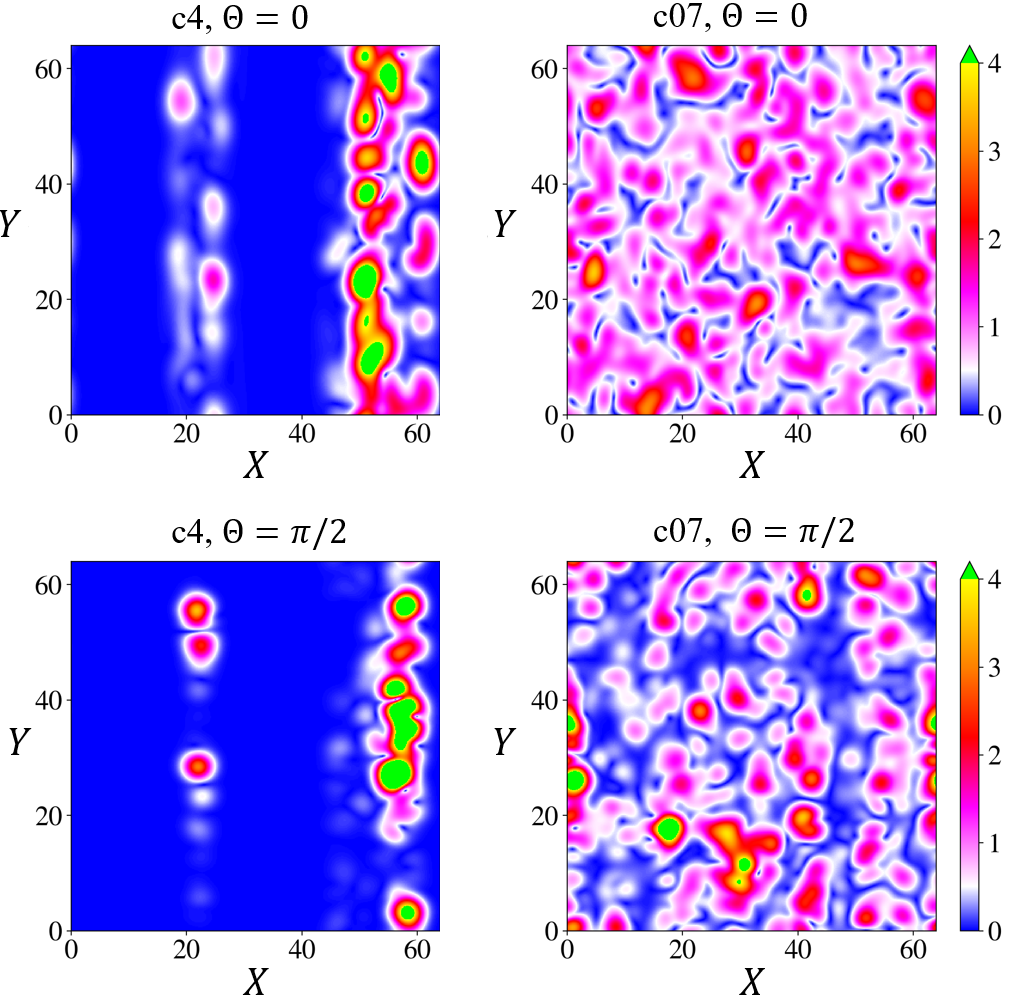}
\end{center}
\caption{Square root of the LIM, $[I(k,{\bm X},\Theta)]^{1/2}$, at $k \approx 3.9$ for $\Theta=0$ and $\pi/2$.} 
\label{visW2D}
\end{figure}
The CWT is applied to the vorticity field $\omega$ at time $t_s$.
The field is shown in Fig.~\ref{viFLD}. 
We consider the local wavelet spectrum $|W[\omega](k,{\bm X},\Theta)|^2/a^2$. 
Here, ${\bm X}=(X,Y)$ denotes the position, and the scale $a$ is approximately related to the wavenumber $k$ through Eq.~(\ref{ak}).
For visualization, we average the local wavelet spectrum over $Y$ and denote the $Y$-averaged quantity by $\left\langle |W[\omega](k,{\bm X},\Theta)|^2 \right\rangle_Y/a^2$.
This averaging reduces the influence of local fluctuations in the $y$-direction and gives an $(X,k)$ representation without choosing a particular value of $Y$.
Figure~\ref{visW1D} shows the scalograms for c4 and c07 at two representative angles, $\Theta=0$ and $\pi/2$.
It can be seen that, for c4 at $\Theta=0$, high- and low-intensity regions at wavenumbers $k \gtrsim 2$, corresponding to small scales, are associated with more turbulent and much less turbulent regions in Fig.~\ref{viFLD}, respectively.
By contrast, for c07 at $\Theta=0$, high-intensity regions are seen over the whole domain, together with a stripe-like pattern at wavenumbers $k \gtrsim 2$. 
This pattern is in accordance with the spatial structure shown in Fig.~\ref{viFLD}. 
For both c4 and c07, the intensity appears stronger for $\Theta=0$ than for $\Theta=\pi/2$, 
and the high-intensity regions at small scales look more localized for $\Theta=\pi/2$ than for $\Theta=0$. 
This suggests that spatial intermittency is stronger for $\Theta=\pi/2$ than for $\Theta=0$.
Here, we consider the range $0.79 \lesssim k \lesssim 15.7$, which corresponds to $0.006 \lesssim a/L \lesssim 0.12$. 
Thus, the scales considered here are much smaller than $L(=64)$. 
The upper end of this $k$ range lies sufficiently below the maximum wavenumber retained after dealiasing. 

Figure~\ref{visW2D} presents the square root of the LIM, $[I(k,{\bm X},\Theta)]^{1/2}$, in the $(X,Y)$ plane for $\Theta=0$ and $\pi/2$ at a wavenumber $k\approx 3.9$. 
Here, $k\approx3.9$ is selected because the spatial features described above are visible in the scalograms in Fig.~\ref{visW1D} at this wavenumber.
The LIM $I$ is defined by Eq.~(\ref{LIM}).
For c4, regions of large $I^{1/2}$ are associated with more turbulent regions in Fig.~\ref{viFLD}. 
Within these regions, the distribution for $\Theta=\pi/2$ looks sparser than that for $\Theta=0$. 
For c07 at $\Theta=0$, regions of intermediate values, $0.5 \lesssim I^{1/2} \lesssim 1.5$, are seen over much of the domain, whereas regions of larger values are sparsely distributed. 
For $\Theta=\pi/2$, the regions of larger values appear more spatially localized than those for $\Theta=0$.
\subsection{Wavelet statistics}
We next examine the statistical properties of the wavelet coefficients.
The wavelet statistics are obtained from 75 snapshots sampled every 2 time units, as described in Sec.~\ref{subsec:DNS}.

\begin{figure}[bt!]
\begin{center}
\includegraphics[width=7.5cm,keepaspectratio]
{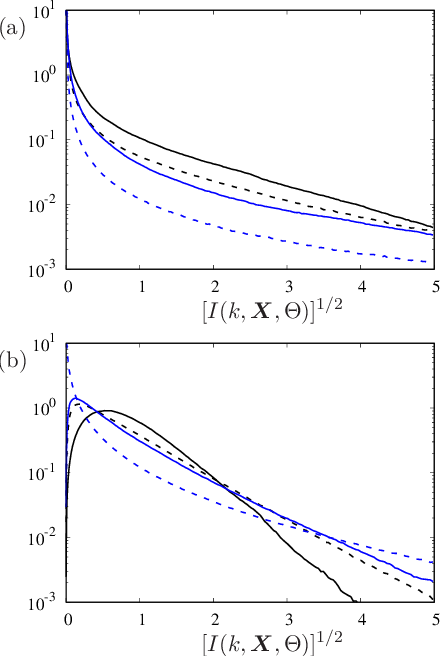}
\end{center}
\caption{PDFs of $[I(k,{\bm X},\Theta)]^{1/2}$ for $\Theta=0$ and $\pi/2$ at $k\approx 3.9$ and $7.9$: (a) c4 and (b) c07.
Solid and dashed curves represent $\Theta=0$ and $\pi/2$, respectively; black and blue curves represent $k\approx 3.9$ and $7.9$, respectively.
}\label{figPDF}
\end{figure}
Figure~\ref{figPDF} shows the probability density functions (PDFs) of $[I(k,{\bm X},\Theta)]^{1/2}$ for $\Theta=0$ and $\pi/2$ at the two wavenumbers $k\approx 3.9$ and $7.9$.
The second wavenumber, $k\approx7.9$, is approximately twice $k\approx3.9$ and is used to examine the statistics at a smaller scale.
In the following discussion, the comparisons refer to the range shown in Fig.~\ref{figPDF}.
In c4, at $k\approx 3.9$, the PDFs for both $\Theta=0$ and $\pi/2$ take high values near $I^{1/2}=0$.
At $k\approx 7.9$, this near-zero part of the PDFs becomes more pronounced.
For each wavenumber, the tail for $\Theta=0$ lies above the tail for $\Theta=\pi/2$.
For each angle, the PDF at large $I^{1/2}$ is higher for $k\approx 3.9$ than for $k\approx 7.9$.
In c07, at each wavenumber, the PDF for $\Theta=0$ exceeds the PDF for $\Theta=\pi/2$ over an intermediate range of $I^{1/2}$.
By contrast, the PDF for $\Theta=\pi/2$ is larger than that for $\Theta=0$ near $I^{1/2}=0$ and at large $I^{1/2}$.
For each angle, the PDF at $k\approx 3.9$ exceeds the PDF at $k\approx 7.9$ over an intermediate range of $I^{1/2}$, whereas the PDF at $k\approx 7.9$ is larger near $I^{1/2}=0$ and at large $I^{1/2}$.
The features of the PDFs at $k\approx 3.9$ are in accordance with the spatial distributions of $[I(k,{\bm X},\Theta)]^{1/2}$ shown in Fig.~\ref{visW2D}, as discussed in Sec.~\ref{Viswave}.

\begin{figure}[bt!]
\begin{center}
\includegraphics[width=7.5cm,keepaspectratio]{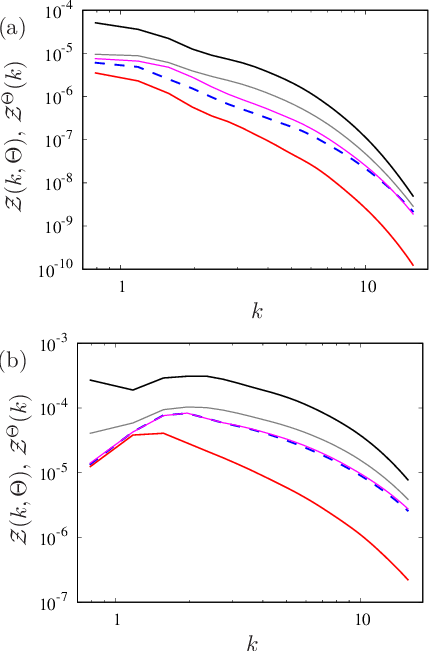}
\end{center}
\caption{The global wavelet spectra ${\mathcal Z}(k,\Theta)$ vs. $k$ for (a) c4 and (b) c07. 
The black, blue, red, and magenta curves denote ${\mathcal Z}(k,\Theta)$ at $\Theta=0$, $\pi/4$, $\pi/2$, and $3\pi/4$, respectively. 
The gray curves denote the angle-averaged spectrum ${\mathcal Z}^\Theta(k)$.
} \label{figWS}
\end{figure}
\begin{figure}[bt!]
\begin{center}
\includegraphics[width=7.5cm,keepaspectratio]{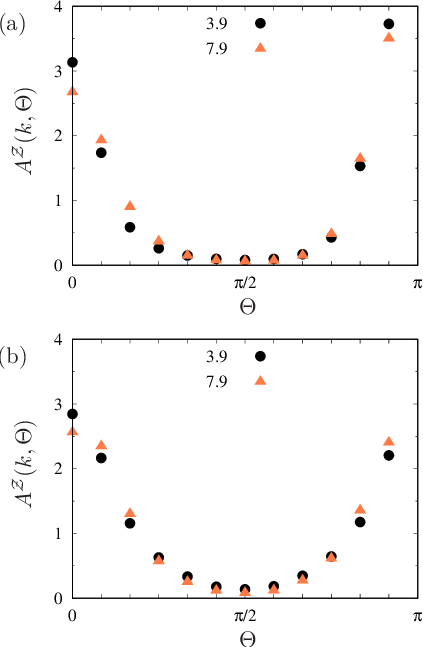}
\end{center}
\caption{Scale-dependent anisotropy measure $A^{\mathcal Z}(k,\Theta)$ vs. $\Theta$ at $k\approx 3.9$ and $7.9$ for (a) c4 and (b) c07.} \label{figWSani}
\end{figure}

The following global statistics are used to characterize small-scale tendencies of the flow in the presence of zonal flows.
Figure~\ref{figWS} shows the global wavelet spectra of vorticity $\omega$, ${\mathcal Z}(k,\Theta)$, defined by Eq.~(\ref{WSpe}). 
These spectra characterize the directional anisotropy of $\omega$.
For c4 and c07, ${\mathcal Z}(k,\Theta)$ decreases with increasing $k$ for $k \gtrsim 2$. 
At a fixed $k$ $(\gtrsim 2)$, ${\mathcal Z}(k,0)$ is larger than ${\mathcal Z}(k,\Theta)$ at the other angles shown, whereas ${\mathcal Z}(k,\pi/2)$ is smaller. 
The angle-averaged spectrum ${\mathcal Z}^\Theta(k)$ defined by Eq.~(\ref{AveWSpe}) is also shown for reference. 
The range $0 \le \Theta < \pi$ is discretized into 12 equally spaced angles with spacing $\pi/6$, 
as in Ref.~\onlinecite{KaiPRX}, in which a 2D Morlet wavelet was also used.

Figure~\ref{figWSani} shows the scale-dependent anisotropy measure $A^{\mathcal Z}(k,\Theta)$ defined by Eq.~(\ref{AniE}) at $k\approx 3.9$ and $7.9$.
Since the statistics are based on the modulus of the wavelet coefficient, $\Theta=0$ and $\Theta=\pi$ are equivalent.
For both c4 and c07, $A^{\mathcal Z}(k,\Theta)$ takes large values around $\Theta=0$ and small values around $\Theta=\pi/2$ at the two wavenumbers.
The profiles of $A^{\mathcal Z}(k,\Theta)$ appear nearly symmetric about $\Theta=\pi/2$, particularly for c07.
Thus, the global wavelet spectra are enhanced around directions perpendicular to the zonal flow.
The enhancement around $\Theta=0$ is consistent with the stripe-like structures seen in physical space, because $\Theta=0$ corresponds to local wave vectors perpendicular to the zonal-flow direction.

In the MHW model defined by Eqs.~(\ref{MHWome}) and (\ref{MHWden}), the mean density gradient and the zonal flow define distinguished directions along the $x$- and $y$-axes, respectively. 
In this work, we consider wavelet statistics for $0 \le \Theta < \pi$, owing to the $\pi$-periodicity of the modulus of the wavelet coefficient.
This makes symmetry about $\Theta=\pi/2$ a natural reference for interpreting the directional statistics.
In numerical simulations, however, the small-scale statistics may be influenced by organized flow structures, which can lead to deviations from this symmetry.

\begin{figure}[bt!]
\begin{center}
\includegraphics[width=7.5cm,keepaspectratio]{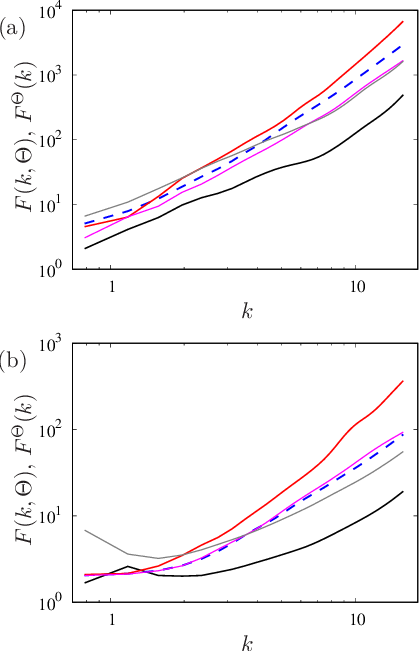}
\end{center}
\caption{Wavelet flatness $F(k,\Theta)$ vs. $k$ for (a) c4 and (b) c07. 
The black, blue, red, and magenta curves denote $F(k,\Theta)$ at $\Theta=0$, $\pi/4$, $\pi/2$, and $3\pi/4$, respectively. 
The gray curves denote the angle-averaged flatness $F^\Theta(k)$.} \label{figWF}
\end{figure}
\begin{figure}[bt!]
\begin{center}
\includegraphics[width=7.5cm,keepaspectratio]{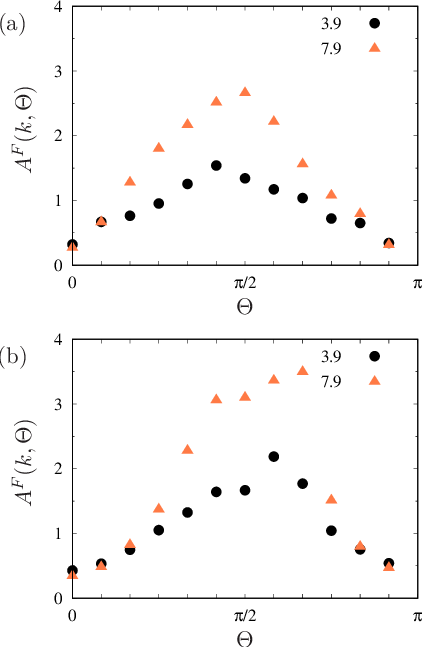}
\end{center}
\caption{Scale-dependent anisotropy measure $A^F(k,\Theta)$ vs. $\Theta$ at $k \approx 3.9$ and $7.9$ for (a) c4 and (b) c07.} \label{figWFani}
\end{figure}

Figure~\ref{figWF} shows the wavelet flatness of $\omega$, $F(k,\Theta)$, defined by Eq.~(\ref{WFlat}).
For c4 and c07, $F(k,\Theta)$ increases with increasing $k$ for $k \gtrsim 2$, which is characteristic of spatial intermittency.
In this range, the flatness values are larger for c4 than for c07.
At a fixed $k \gtrsim 2$, among the angles shown, $F(k,\pi/2)$ tends to take larger values, whereas $F(k,0)$ tends to take smaller values.
This enhancement reflects larger contributions of high LIM values in Eq.~(\ref{FlatLIM}).
Figure~\ref{figWF} also shows $F^\Theta(k)$, defined by Eq.~(\ref{aveFlat}) from the angle-averaged second- and fourth-order moments, for reference.
Figure~\ref{figWFani} presents the scale-dependent anisotropy measure $A^F(k,\Theta)$ defined by Eq.~(\ref{AniF}) at $k \approx 3.9$ and $7.9$.
The values of $A^F(k,\Theta)$ tend to be larger around $\Theta=\pi/2$ than around $\Theta=0$, suggesting that directional intermittency is generally stronger around directions along the zonal flow.
The profiles of $A^F(k,\Theta)$ appear less symmetric about $\Theta=\pi/2$ than those of $A^{\mathcal Z}(k,\Theta)$.
In most cases, the maxima of $A^F(k,\Theta)$ are not located at $\Theta=\pi/2$.
This may also reflect the influence of organized flow structures.
The resulting deviations may be more pronounced in fourth-order statistics than in second-order statistics.
Fourth-order statistics are more sensitive to intense localized events.
The statistics shown here are obtained from the 75 snapshots described above.

The angular dependence observed here may be compared with that reported for three-dimensional homogeneous turbulence with uniform stable density stratification.
The Fourier energy spectra are larger along the mean density-gradient direction than in the perpendicular plane (see, e.g., Refs.~\onlinecite{GODEFERDSTAQUET2003,SagautCambon2018}).
The flatness based on orthogonal wavelet coefficients is much larger in directions perpendicular to the mean density gradient than in those parallel to it.\cite{BLS2007}

Wavelet spectra can also be conditioned on spatial regions.
The corresponding spectra are shown in Appendix~\ref{appen}.

\section{Conclusions}\label{sec:concl}
We have applied a 2D directional continuous wavelet transform to vorticity fields of drift-wave turbulence with zonal flows.
The fields were obtained from numerical simulations of the modified Hasegawa--Wakatani model.
The CWT with a Morlet wavelet has been used to characterize the turbulence in terms of scale, position, and direction.
We have examined small-scale anisotropy and spatial intermittency by wavelet statistics based on the second- and fourth-order moments.
It has been shown that the wavelet spectra are anisotropic and take larger values around directions perpendicular to the zonal flow.
The wavelet flatness is larger at smaller scales. 
This increase toward smaller scales reflects stronger spatial localization of intense fluctuations.
Its angular dependence shows that spatial intermittency is generally stronger around directions along the zonal flow.

This analysis is also applicable to other fields, such as the density fluctuation $n$ and electrostatic potential $\phi$.
For fields with weaker small-scale fluctuations, however, the wavenumber range that can be examined numerically may not extend as far toward high wavenumbers as for vorticity.
In the present study, the CWT has been applied only in space, and the resulting wavelet statistics have been examined over the statistically quasi-stationary period.
A direct analysis of features such as temporal intermittency and burst-like events is therefore outside the scope of this work.
A localized spatio-temporal description of plasma turbulence remains to be developed by extending the present framework.
Such an extension would provide deeper insight into transport and energy transfer in the presence of coupling between drift waves and zonal flows.
Ensemble averaging over multiple realizations will also be important for developing such a description in a more systematic way.
It would be useful to decompose the flow into zonal and non-zonal components prior to such spatio-temporal analysis.
\acknowledgments

This work used the computational resources of the Plasma Simulator NEC SX-Aurora TSUBASA at the National Institute for Fusion Science (NIFS), and the Plasma Simulator NEC LX 204Bin-3 jointly operated by NIFS and the National Institutes for Quantum Science and Technology.
This study was carried out with the support and under the auspices of the NIFS Collaboration Research program (NIFS25KISC024).
KY was also supported by the JSPS KAKENHI Grant Number JP25K07151.

ZL and KS acknowledge funding from the French Federation for Magnetic Fusion Studies (FR-FCM) and the EUROfusion Consortium, funded by the Euratom Research and Training Programme under Grant Agreement No.~633053. 
The views and opinions expressed herein do not necessarily reflect those of the European Commission. 
ZL and KS also acknowledge support from the Institute for Fusion Sciences and Instrumentation in Nuclear Environments (ISFIN) at Aix-Marseille Universit\'e, 
funded by the French government under the France 2030 program managed by the A$^*$MIDEX initiative (AMX-19-IET-013). 
The authors were also granted access to the HPC resources of the Centre de Calcul Intensif d’Aix-Marseille Universit\'e where the numerical simulations were carried out.
The authors thank Benjamin Kadoch for providing the numerical simulation code.

\appendix
\section{Conditioned wavelet spectra}\label{appen}
\begin{figure}[bt!]
\begin{center}
\includegraphics[width=7.5cm,keepaspectratio]{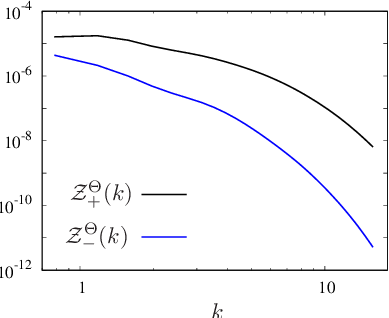}
\end{center}
\caption{Angle-averaged conditioned wavelet spectra ${\mathcal Z}_\pm^\Theta(k)$ vs. $k$.
Here, ${\mathcal Z}_+^\Theta(k)$ and ${\mathcal Z}_-^\Theta(k)$ correspond to the regions with $U_y>0$ and $U_y<0$, respectively.} \label{figWS_C}
\end{figure}
We briefly define the conditioned wavelet spectra and present the corresponding results for c4.
The conditioning is imposed according to the sign of the zonal-flow velocity $U_y(X,t)$ at each time.
We define the subdomains $\Omega_{\pm}(t)$ by $ \Omega_{\pm}(t)=\left\{\bm{X}\;\middle|\;U_y(X,t)\gtrless 0\right\}.$
We focus on the conditioned second-order statistics given by $M_{2,\pm}(k,\Theta) =\left\langle|W[\omega](k,\bm{X},\Theta)|^2 \right\rangle_{\pm}$,
where $\langle  \cdot \rangle_{\pm}$ denotes the average over $\Omega_{\pm} (t)$ and over time.
As an example, we present the conditioned angle-averaged wavelet spectrum defined by
${\mathcal Z}^\Theta_\pm(k)=\left\langle M_{2,\pm}(k,\Theta)\right\rangle_\Theta/a^2$. 
At the time shown for c4 in Fig.~\ref{viFLD}, $\Omega_{+}$ and $\Omega_{-}$ correspond to the more turbulent and much less turbulent regions, respectively.
Figure~\ref{figWS_C} shows that ${\mathcal Z}_{+}^\Theta(k)$ is larger than ${\mathcal Z}_{-}^\Theta(k)$,
while ${\mathcal Z}_{-}^\Theta(k)$ still exhibits multiscale features, suggesting that fluctuations remain over a range of scales even in the weakly fluctuating region.

\bibliographystyle{apsrev4-1}
\bibliography{ref}
\end{document}